\documentclass[reprint,superscriptaddress,amsmath,amssymb,prb,longbibliography]{revtex4-1}
\usepackage{comment}
\usepackage{color}
\usepackage{graphicx}
\usepackage{subfigure}
\usepackage{dcolumn}
\usepackage{bm}
\usepackage{amsmath}
 
\usepackage{braket}
\usepackage{diagbox, eqparbox, hhline}
\usepackage[detect-none]{siunitx}
\DeclareMathAlphabet{\mathpzc}{OT1}{pzc}{m}{it}

\newcommand{\Keff}{K_{\text{eff}}} 
\newcommand{\Msat}{M_{\text{s}}}
\newcommand{\Fmax}{F_{\text{max}}}

\begin{document}


\title{Deflection of (anti)ferromagnetic skyrmions at heterochiral interfaces}
\author{Ra\'i M. Menezes}
\affiliation{%
Departement Fysica, Universiteit Antwerpen, Groenenborgerlaan 171, B-2020 Antwerpen, Belgium
}
\affiliation{%
Departamento de F\'isica, Universidade Federal de Pernambuco, Cidade Universit\'aria, 50670-901, Recife-PE, Brazil
}%
\author{Jeroen Mulkers}
\affiliation{%
Departement Fysica, Universiteit Antwerpen, Groenenborgerlaan 171, B-2020 Antwerpen, Belgium
}
\affiliation{%
DyNaMat Lab, Department of Solid State Sciences, Ghent University, B-9000 Ghent, Belgium
}%
\author{Cl\'ecio C. de Souza Silva}
\affiliation{%
Departamento de F\'isica, Universidade Federal de Pernambuco, Cidade Universit\'aria, 50670-901, Recife-PE, Brazil
}%
\author{Milorad V. Milo\v{s}evi\'c}%
\email{milorad.milosevic@uantwerpen.be}
\affiliation{%
Departement Fysica, Universiteit Antwerpen, Groenenborgerlaan 171, B-2020 Antwerpen, Belgium
}%

\date{\today}

\begin{abstract}
Devising magnetic nanostructures with spatially heterogeneous Dzyaloshinskii-Moriya interaction (DMI) is a promising pathway towards advanced confinement and control of magnetic skyrmions in potential devices. Here we discuss theoretically how a skyrmion interacts with a heterochiral interface using micromagnetic simulations and analytic arguments. We show that a heterochiral interface deflects the trajectory of ferromagnetic (FM) skyrmions, and that the extent of such deflection is tuned by the applied spin-polarized current and the difference in DMI across the interface. Further, we show that this deflection is characteristic for the FM skyrmion, and is completely absent in the antiferromagnetic (AFM) case. In turn, we reveal that the AFM skyrmion achieves much higher velocities than its FM counterpart, yet experiences far stronger confinement in nanoengineered heterochiral tracks, which reinforces AFM skyrmions as a favorable choice for skyrmion-based devices.  
\end{abstract}

\pacs{Valid PACS appear here}
\maketitle

\section{Introduction}

The interfacially induced Dzyaloshinskii-Moriya interaction (DMI) is a chiral interaction observed in ferromagnetic thin films, e.g., a Co layer, when coupled to nonmagnetic layers with a strong spin-orbit coupling, e.g., the heavy metal Pt \cite{Dzyaloshinsky1958AAntiferromagnetics,Moriya1960AnisotropicFerromagnetism,Crepieux1998Dzyaloshinsky-MoriyaSurface}. The DMI favors the rotation of the magnetization at short length scales, giving rise to  chiral spin structures such as cycloids and magnetic skyrmions \cite{Bogdanov1994ThermodynamicallyCrystals,Bogdanov2001ChiralMultilayers,Ezawa2011CompactFilms,Kiselev2011ChiralTechnologies}. Particularly, magnetic skyrmions are promising candidates for technological applications, such as spin-based information processing and computing devices \cite{ParkinMagneticMemory,Fert2013SkyrmionsTrack,Kiselev2011ChiralTechnologies,Nagaosa2013TopologicalSkyrmions}. Most recently, the suggestion of skyrmions in antiferromagnetic (AFM) systems has also increased the expectation on skyrmion-based devices, since in those systems the skyrmions are not sensitive to stray fields, move straight along the direction imposed by the applied current, and present better mobility with lower energy costs \cite{Barker2016StaticTemperature,Zhang2016AntiferromagneticManipulation,Xia2017ControlRacetrack,Jin2016DynamicsEffect,rosales2015three,shen2018dynamics}.

The confinement of ferromagnetic (FM) skyrmions in mesoscopic chiral films and tracks has already been thoroughly studied in recent years \cite{Rohart2013SkyrmionInteraction,Mulkers2016CycloidalMagnets,Chui2015GeometricalNanowire,Navau2016AnalyticalNano-oscillators}. As a latest development, spatial engineering of DMI has been suggested as an alternative manner of skyrmion guidance and manipulation. Such heterochiral samples have been demonstrated to strongly confine magnetic skyrmions\cite{Mulkers2017EffectsFilms}, pin them or manipulate their size\cite{Stosic2017pinning}, and increase their lifetime\cite{Stosic2017PathsFilms}, in the regions where the DMI is higher. The interest in these results is reinforced by the fact that heterochiral structures can indeed be fabricated experimentally, via engineering of the substrate and/or the capping layer of the thin ferromagnetic film \cite{Balk2017SimultaneousTrilayers,Wells2017EffectPt/Co/Pt}. Bearing in mind the potential of heterochiral systems for the development of skyrmion-based devices, the last piece of the puzzle is to understand the skyrmion dynamics in such samples. However, the dynamics of a single magnetic skyrmion while e.g. crossing the regions with different DMI strengths remains mostly unexplored, while the confinement effects in AFM heterochiral films have not been studied at all to date. 

\begin{figure}[!t]
\centering
\includegraphics[width=\linewidth]{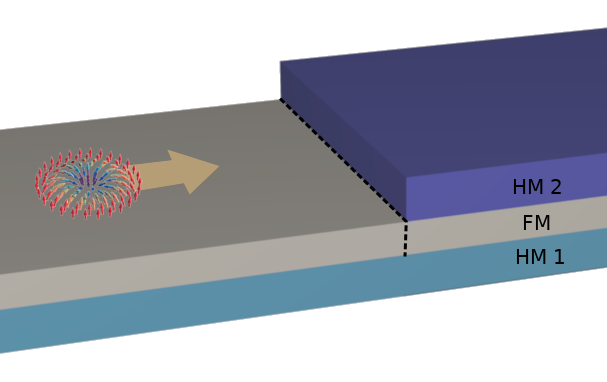}
\caption{Schematic representation of an experimental analogue of the considered system, a ferromagnetic layer between two heavy metal (HM) layers, with a suitably patterned top layer. In the depicted sample, the heterochiral interface is created at the lateral end of the top layer. The dashed line indicates the interface where the DMI changes.}
\label{Fig1}
\end{figure}

Therefore, in this work we address theoretically the dynamics of FM and AFM skyrmions in heterochiral films, and particularly their interaction with a heterochiral interface (where DMI changes, see e.g. a suggested realization in Fig. \ref{Fig1}). The DMI strength can be modified along the sample using, e.g., lithographic techniques to correspondingly pattern the heavy metal (HM) layer(s) and thereby adjust the interfacially-induced DMI\cite{Mulkers2017EffectsFilms,Balk2017SimultaneousTrilayers,Wells2017EffectPt/Co/Pt}. Note that by altering the HM configuration and/or thickness one might also induce changes in other material parameters, such as the magnetic anisotropy. In this paper, we are focusing exclusively on the effects of the spatially varied DMI, and therefore consider the other material parameters homogeneous throughout the sample to avoid ill interpretations of the results. We proceed by employing micromagnetic simulations to show that local canting of the magnetization at the heterochiral interface\cite{Mulkers2017EffectsFilms} can be seen as an imposed potential barrier in the Thiele formalism for the center-of-mass of the skyrmion, that causes a characteristic deflection in the trajectory of the FM skyrmion when crossing the heterochiral interface. After verifying it in full micromagnetic simulations, we show that such deflection is completely absent in an analogous antiferromagnetic sample, and that the AFM skyrmion: (i) moves much faster than the FM skyrmion, as already predicted in the literature\cite{Barker2016StaticTemperature}, but (ii) experiences far stronger confinement in heterochiral films, so that the critical current needed to push it over a heterochiral interface is much larger than in the FM case. These results promote antiferromagnetic heterochiral films as an advanced platform for skyrmion-based devices.

The paper is organized as follows. In Sec.~\ref{Sec.II} we describe the micromagnetic model of ferromagnetic films with interfacially induced DMI and applied spin-polarized current, before providing some analytic considerations and the Thiele formalism for the center-of-mass motion of the FM skyrmion driven by in-plane and out-of-plane spin-polarized currents. In Sec.~\ref{Sec.III}, we report the characteristic features of skyrmion motion when crossing an interface where DMI changes, for both the ferromagnetic (Sec.~\ref{Sec.IIIA}) and antiferromagnetic (Sec.~\ref{Sec.IIIB}) cases, and analyze them using both micromagnetic simulations and the Thiele derivations. In Sec.~\ref{Sec.IIIC} we show a possible application of heterochiral interfaces for the manipulation of a skyrmion chain in a nanoengineered circuit. Finally, our results are summarized in Sec.~\ref{Sec.IV}.  

\section{Theoretical formalism}\label{Sec.II}

\subsection{Micromagnetic model}

For the micromagnetic simulations, we employ the micromagnetic simulation package mumax$^3$ ~\cite{Vansteenkiste2014TheMuMax3}, on an ultrathin ferromagnetic film with perpendicular magnetic anisotropy and with spatially inhomogeneous DMI. The local free energy density $\mathcal{E}$, related to the magnetization $\Vec{M}(x,y)=\Msat\Vec{m}(x,y)$, has multiple sources, and we consider the following: exchange, anisotropy, DMI, and demagnetization. We approximate the demagnetization energy by using an effective anisotropy $\Keff=K-\frac{1}{2}\mu_0 \Msat^2$. This approximation is justified by the fact that we are interested in ultrathin films, where dipolar coupling becomes local in the zero-thickness limit \cite{coey2010magnetism}. In this work we do not consider the effects of external magnetic field, and the Zeeman term of the energy-density is therefore zero. The expressions for the remaining energy-density terms are
\begin{subequations}\nonumber
\begin{align}
    &\mathcal{E}_{\text{ex}}=A\left[ (\partial_x\textbf{m})^2 +(\partial_y\textbf{m})^2\right],\\
    &\mathcal{E}_{\text{anis}}=K_{\text{eff}}(1-m_z^2),\\
    &\mathcal{E}_{\text{dmi}}=D\left[m_x\partial_x m_z-m_z\partial_x m_x
    +m_y\partial_y m_z -m_z\partial_y m_y\right].
\end{align}
\end{subequations}
For the simulations of the ferromagnetic case, we consider the following parameters:
saturation magnetization: $\Msat=580$~kAm$^{-1}$, exchange stiffness: $A=15$~pJm$^{-1}$, and perpendicular anisotropy $K=0.8 $~MJm$^{-3}$ ($\Keff=0.6$~MJm$^{-3}$). The used values of the DMI constant, $D$, will be specified in the sections below. For all simulations, we consider a system discretized into cells of size $1\times1\times0.4$ nm$^3$. The dynamics of the magnetization is governed by the Landau-Lifshitz-Gilbert (LLG) equation
\begin{equation}
    \frac{d\textbf{m}}{dt}=\frac{\gamma}{1+\alpha^2}\left(\textbf{m}\times\textbf{H}_{\text{eff}}+\alpha\left[\textbf{m}\times(\textbf{m}\times\textbf{H}_{\text{eff}})\right]\right), 
\end{equation} 
where $\gamma$ is the gyromagnetic ratio and $\alpha$ the damping factor. $\textbf{H}_{\text{eff}}$ is the effective magnetic field given by the functional derivative of the free energy $E=\int\mathcal{E}dV$ with respect to the magnetization: $\textbf{H}_{\text{eff}}=-\frac{1}{\mu_0 \Msat}\delta E/\delta\textbf{m}$. 

The N\'eel skyrmion in a chiral magnetic film can be driven by two different scenarios \cite{Tomasello2014AMemories}: (i) by an in-plane spin-polarized current (CIP) applied into the magnetic layer, or (ii) by an electrical current applied into the HM layer, which due to the spin Hall effect gives rise to a spin-polarized current perpendicular to the plane (CPP)\cite{Tomasello2014AMemories,Jin2016DynamicsEffect,Zhang2016MagneticEffect,Jiang2017DirectEffect}. In this work we explore both scenarios. For simulations of the spin transfer torque (STT) associated with the CIP and CPP scenarios, the Zhang and Li \cite{Zhang2004RolesFerromagnets} and Slonczewski \cite{Slonczewski1996Current-drivenMultilayers,Xiao2004BoltzmannTorque} STT terms, respectively, were added to the LLG equation. Both kinds of STT are implemented in the simulation package mumax$^3$. In the CPP scenario, the electrical current applied into the HM layer results in a spin current injected into the FM along the $z$ direction, with  $\textbf{m}_p=-\text{sgn}\theta_{SH}(\hat{z}\times\hat{j}_{\text{hm}})$ the orientation  of the injected spins \cite{Tomasello2014AMemories,liu2012current,perez2014micromagnetic,Jin2016DynamicsEffect}, where $\theta_{SH}$ is the spin-Hall angle of the HM. In mumax$^3$ one can simulate similar scenario by considering a fixed layer, with polarization vector $\textbf{m}_p$, on top of the film and the applied current injected along the $\hat{z}$ direction. For the CPP simulations we consider the Slonczewski parameter $\Lambda=1$, and secondary spin-torque parameter $\epsilon'=0$ (for details of equations we refer to Ref. \onlinecite{Vansteenkiste2014TheMuMax3}). For both the CIP and the CPP scenario, the polarization rate of the spin-polarized current is fixed at $P=0.4$. The other relevant parameters associated to the STT, such as the non-adiabaticity factor $\beta$ (CIP scenario) and the fixed-layer polarization (CPP scenario), will be specified in the next sections.

For the antiferromagnetic samples, we consider the same parameters of the FM case, except for the negative exchange stiffness $A=-15$ pJm$^{-1}$. Note that mumax$^3$ was originally developed for simulations of FM systems in the continuous field approximation. However, once we consider the AFM system, which comprises two sublattices of reversely-aligned spins, we end up performing an atomistic simulation (albeit with a large lattice parameter), where the finite-differences derivatives performed by mumax$^3$ are mathematically equivalent to the classical Heisenberg model \cite{Vansteenkiste2014TheMuMax3}. The STT can be applied also to the AFM system provided one considers an ultra-small mesh size in the micromagnetic simulations\cite{Gomonay2010SpinAntiferromagnets,Gomonay2012SymmetryCurrent,Zhang2016AntiferromagneticManipulation,Jin2016DynamicsEffect,Xia2017ControlRacetrack}. In this work, we simulate only the CPP-driven AFM skyrmion. Note that one can not straightforwardly use the CIP scenario in the micromagnetic simulations since spatial derivatives are involved in the STT term and the reversely-aligned magnetization of the AFM system can no longer be described by a differentiable field.

\subsection{Thiele equation for skyrmion dynamics}
The Thiele equation describes the dynamics of the center-of-mass of the skyrmion by assuming a rigid body motion of the spin texture \cite{Tomasello2014AMemories,Zhang2016MagneticEffect,Jiang2017DirectEffect,Iwasaki2014ColossalEdge}, and can be written out for both CIP and CPP scenarios. 

The Thiele equation for the CIP scenario reads 
\begin{equation}
    \textbf{G}\times(\bm{\nu}-\Dot{\textbf{r}})+\mathcal{D}(\beta\bm{\nu}-\alpha\Dot{\textbf{r}})-\nabla V(\textbf{r})=0, \label{eq.R1}
\end{equation}
where $\textbf{G}=\mathcal{G}\hat{z}=4\pi Q\hat{z}$ is the gyromagnetic coupling vector, with $Q$ the skyrmion number; $\Dot{\textbf{r}}=\Dot{x}\hat{x}+\Dot{y}\hat{y}$ is the drift velocity; $\bm{\nu}$ is the velocity of the conduction electrons associated to the spin-polarized current; $V$ is the potential stemming from an external force, boundaries or impurities; $\mathcal{D}$ represents the dissipative force (for the range of parameters considered in this work, $\mathcal{D}\approx\numrange{4\pi}{8\pi}$). If one considers a current applied along the $x$ direction, i.e. $\nu_y=0$, then the Thiele equation can be separated into its two components, which for the case of $V=0$ yields %
\begin{subequations}
\begin{align}
   \Dot{x}=\left(\frac{\mathcal{G}^2+\mathcal{D}^2\alpha\beta}{\mathcal{G}^2+\mathcal{D}^2\alpha^2} \right)\nu_x, \label{eq.R3a} \\
   \Dot{y}=\left(\mathcal{G}\mathcal{D}\frac{\alpha-\beta}{\mathcal{G}^2+\alpha^2\mathcal{D}^2} \right)\nu_x. \label{eq.R3b}
\end{align}
\end{subequations}
The above equations describe the skyrmion velocity due to the applied current in the absence of external forces and impurities, where the skyrmion velocity is constant for a fixed applied current. Notice that the skyrmion undergoes a transverse motion, $\Dot{y}\neq0$ (when $\alpha$ differs from $\beta$), because it carries a non-zero skyrmion number ($\mathcal{G}\neq 0$). If we consider $V(\textbf{r})=V(x)$, the Thiele equation leads to   
\begin{subequations}
\begin{align}
   \Dot{y}=\frac{\mathcal{G}}{\mathcal{D}\alpha}(\nu_x-\Dot{x}), \label{eq.R4a} \\
   \left(\frac{\mathcal{G}^2}{\mathcal{D}\alpha}+\mathcal{D}\beta  \right)\nu_x-\left(\frac{\mathcal{G}^2}{\mathcal{D}\alpha}+\mathcal{D}\alpha  \right)\Dot{x}=\frac{d V}{d x}. \label{eq.R4b}
\end{align}
\end{subequations}
Note that the $x$ component of the skyrmion velocity is directly affected by the external potential and, consequently, the Magnus force (represented by the $\mathcal{G}$ term), which drives the skyrmion along the $y$ direction, is also affected. Indeed, taking the variation $\delta \Dot{y}\equiv\Dot{y}(t+dt)-\Dot{y}(t)$ of Eq. (\ref{eq.R4a}), for a fixed current density,  we obtain 
\begin{equation}
    \delta\Dot{y}=-\frac{\mathcal{G}}{\mathcal{D}\alpha}\delta\Dot{x},
    \label{Eq.x..}
\end{equation}
which means that, if the external potential is attractive (repulsive) in the $x$ direction, the skyrmion trajectory is deflected to the $-\mathcal{G}\hat{y}$ ($+\mathcal{G}\hat{y}$) direction, depending on the skyrmion number.   

In the case of a repulsive external potential, the critical current for the skyrmion to overcome such energy barrier is given by choosing $\Dot{x}=0$ for the maximal value of $F=dV/dx$ in Eq. (\ref{eq.R4b}), i.e. 
\begin{equation}
    \nu_x^c=\frac{\Fmax}{\frac{\mathcal{G}^2}{\mathcal{D}\alpha}+\mathcal{D}\beta  }.
    \label{jcCIP}
\end{equation}
For the limit of small $\alpha$ and $\beta$ ($\alpha\sim\beta\ll1$), the critical current can be approximated as $\nu_x^c=\frac{\Fmax\mathcal{D}\alpha}{\mathcal{G}^2}$. In the same way, Eq. (\ref{eq.R4b}) results in $(\nu_x-\Dot{x})\approx\frac{\mathcal{D}\alpha}{\mathcal{G}^2}\frac{dV}{dx}$. Substituting this expression into Eq. (\ref{eq.R4a}), we obtain the velocity of the skyrmion in the $y$ direction:
\begin{equation}
    \Dot{y}\approx\frac{1}{\mathcal{G}}\frac{dV}{dx}, \label{eq.ymax}
\end{equation}
which depends only on the external potential $V$. The maximal velocity can be written as $\Dot{y}_{\text{max}}=\Fmax/\mathcal{G}$.

Similar results are obtained for a N\'eel skyrmion driven by the CPP scenario. In this case, the skyrmion motion is described by the modified Thiele equation \cite{Tomasello2014AMemories,Zhang2016MagneticEffect,Jiang2017DirectEffect}:
\begin{equation}
    -\textbf{G}\times\Dot{\textbf{r}}-\alpha\mathcal{D}\Dot{\textbf{r}}+4\pi\mathcal{B}\textbf{j}_{\text{hm}}-\nabla V(\textbf{r})=0, \label{eq.CPP}
\end{equation} 
where $\textbf{j}_{\text{hm}}$ is the current density flowing through the heavy metal, which gives rise to a spin-polarized current perpendicular to the plane. The parameter $\mathcal{B}$ quantifies the efficiency of the spin-Hall effect. Now we consider $\textbf{j}_{\text{hm}}=j_{\text{hm}}\hat{y}$. For the case of $V=0$, Eq. (\ref{eq.CPP}) yields
\begin{subequations}
\begin{align}
   \Dot{x}=\frac{\mathcal{G}}{\mathcal{G}^2-\alpha^2\mathcal{D}^2}4\pi\mathcal{B}j_{\text{hm}}, \label{eq.R8a} \\
   \Dot{y}=-\frac{\alpha\mathcal{D}}{\mathcal{G}^2-\alpha^2\mathcal{D}^2} 4\pi\mathcal{B}j_{\text{hm}}. \label{eq.R8b}
\end{align}
\end{subequations}
Note that, for $\alpha\ll1$, the Magnus term dominates $\Dot{x}\gg\Dot{y}$, and the relevant motion is along the $x$ direction.
If we consider $V(\textbf{r})=V(x)$, the modified Thiele equation leads to
\begin{subequations}
\begin{align}
   \Dot{y}=\frac{1}{\mathcal{D}\alpha}(4\pi\mathcal{B}j_{\text{hm}}-\mathcal{G}\Dot{x}), \label{eq.R9a} \\
   -\left(\frac{\mathcal{G}^2}{\mathcal{D}\alpha}+\mathcal{D}\alpha  \right)\Dot{x}+\frac{4\pi\mathcal{B}\mathcal{ G}}{\mathcal{D}\alpha}j_{\text{hm}}=\frac{d V}{d x}. \label{eq.R9b}
\end{align}
\end{subequations}
Taking the variation $ \delta\Dot{y}$ of Eq.~(\ref{eq.R9a}), for a fixed current density,  we recover Eq.~(\ref{Eq.x..}). Therefore, the presence of a external potential deflects the skyrmion trajectory in the same direction as in the CIP scenario. In the same way, the critical current for the skyrmion to overcome a repulsive potential is obtained by choosing $\Dot{x}=0$ in Eq.~(\ref{eq.R9b}), i.e. 
\begin{equation}
    4\pi j_{\text{hm}}^c=\frac{\mathcal{D}\alpha}{\mathcal{G}}\Fmax.
    \label{Eq.R10}
\end{equation}
Finally, for the limit of $\alpha\ll1$, Eq.~(\ref{eq.R9b}) becomes $(4\pi\mathcal{B}j_{\text{hm}}-\mathcal{G}\Dot{x})\approx\frac{\mathcal{D}\alpha}{\mathcal{G}}\frac{d V}{d x}$. By substituting this expression into Eq.~(\ref{eq.R9a}), we recover Eq.~(\ref{eq.ymax}) for the skyrmion velocity in $y$ direction.

\section{Results and discussion}\label{Sec.III}

\subsection{Ferromagnetic skyrmion}\label{Sec.IIIA}

\subsubsection{Skyrmion trajectory when facing nonuniform canting of the background magnetization}\label{Sec.IIIA1}

In this work, we are interested in the skyrmion motion in heterochiral systems, particularly the behavior of the skyrmion trajectory while crossing an interface where DMI changes. As the simplest case of a heterochiral film, we consider a system where the DMI strength, $D$, varies only in the $x$ direction as: $D=D_1$, for $x<x_0$, and $D=D_2$ for $x>x_0$, with $D_1$ and $D_2$ constant. For this geometry, it was shown in Ref.~\onlinecite{Mulkers2017EffectsFilms} that canting of the magnetization is induced at the interface $x=x_0$, and that the magnetization profile is given by:
\begin{equation}
    \theta(x)=2\arctan\left(e^{-|(x-x_0)/\xi|}\tan\frac{\theta_0}{2}\right),\label{eq.theta}
\end{equation}
where $\theta(x)$ is the angle of the spins with respect to the $z$ axis, $m=(\sin\theta,0,\cos\theta)$, $\xi=\sqrt{A/K_{\text{eff}}}$, and 
\begin{equation}
    \theta_0=\arcsin\frac{D_1-D_2}{\pi D_c}
    \label{eq.theta0}
\end{equation}
is the canting angle at the interface, with $D_c=4\sqrt{A\Keff}/\pi$. Notice that the canting of the magnetization at the interface can be either positive or negative, depending on the difference between the DMI strengths $D_1$ and $D_2$.

\begin{figure}[!t]
\centering
\includegraphics[width=\linewidth]{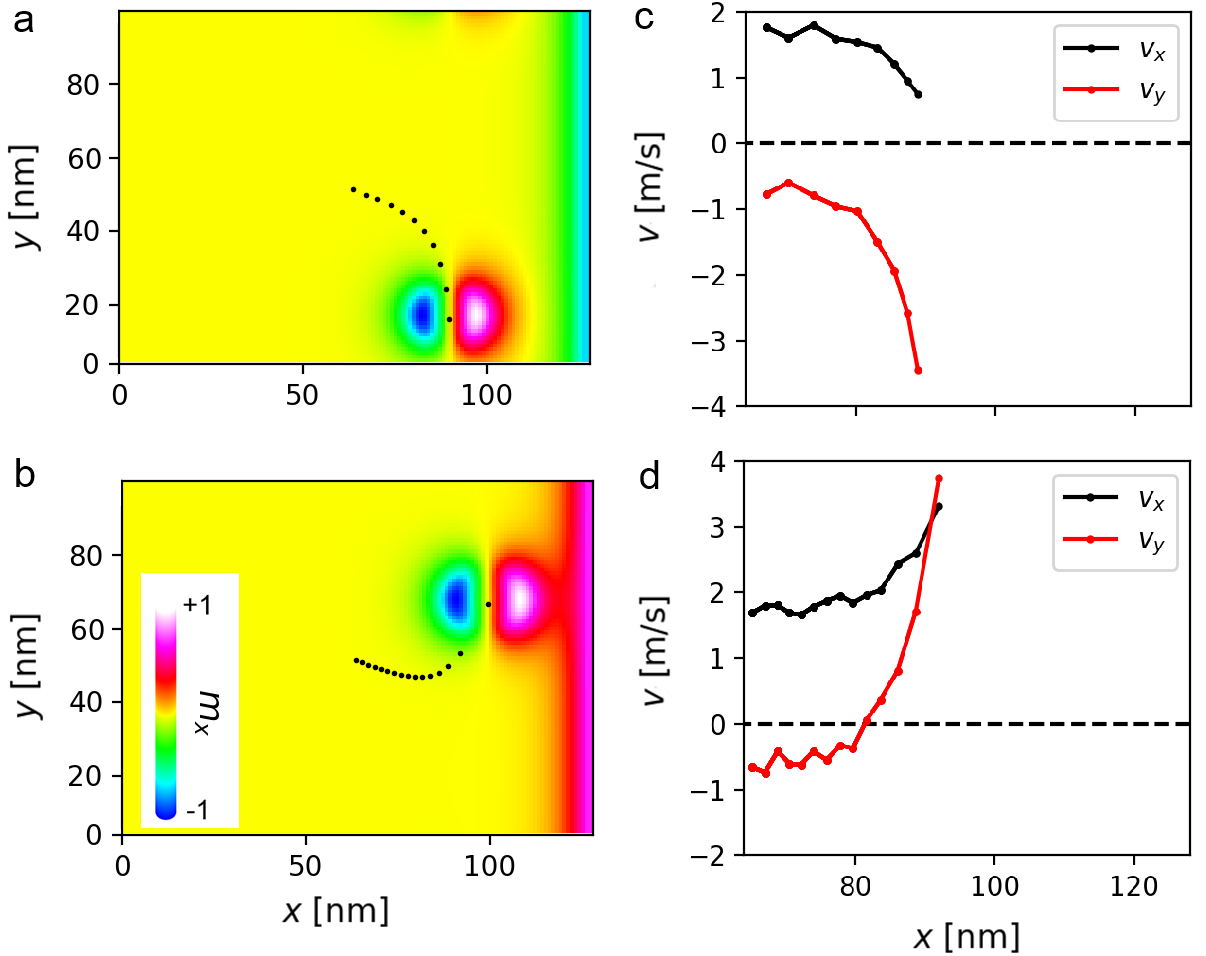}
\caption{Skyrmion trajectories (trail of black dots) in the presence of (a) negative and (b) positive canting of the background magnetization at the right edge of the sample. (c) Center-of-mass velocities of the skyrmion in plot (a). (d) Center-of-mass velocities of the skyrmion in plot (b). Here $j=5\times10^{10}$Am$^{-2}$, $\alpha=0.3$ and $\beta=0$.}
\label{Fig2_3}
\end{figure}

It now becomes of interest to first understand what happens to the skyrmion trajectory when encountering such nonuniform canting of the background magnetization. In Fig. \ref{Fig2_3} we show the result of a simulation performed in the micromagnetic framework for the skyrmion trajectory in a thin ferromagnetic film with uniform DMI, here fixed at $D=0.8D_c$, and artificially imposed canting of spins on the right sample boundary. We consider a sample of size $128\times96\times0.4$ nm$^3$, with periodic boundary conditions in the $y$ direction. The skyrmion is initialized in the center of the simulated region and the energy is minimized numerically. An in-plane polarized current is then applied in the $-\hat{x}$ direction (CIP scenario). The skyrmion undergoes a transverse motion due to the Magnus force. To address the effect of nonuniform canting of the magnetization (as expected at a heterochiral interface), we fix a column of spins at the right side of the sample as $m_{\text{fixed}}=(-\sin\theta_{\text{edge}},0,\cos\theta_{\text{edge}})$ in Fig. \ref{Fig2_3}(a), and $m_{\text{fixed}}=(\sin\theta_{\text{edge}},0,\cos\theta_{\text{edge}})$ in Fig. \ref{Fig2_3}(b), with canting angle $\theta_{\text{edge}}=\frac{\pi}{4}$. The fixed columns spread the canting of the magnetization in the vicinity of the edge, which then affects the skyrmion trajectory. Although these examples are not ideally realistic, they will be useful to understanding the results of the next section. Fig. \ref{Fig2_3}(c,d) shows that the induced canting can be seen as either repulsive (c) or attractive (d) external potential for the skyrmion. In those two cases the Magnus force pushes the skyrmion in different directions ($-\hat{y}$ and $+\hat{y}$). These results are in accordance with Eq. (\ref{Eq.x..}) if one considers the local canting of the magnetization as an external potential for the center of mass of the skyrmion, which is a reasonable assumption, since the energy cost for the skyrmion to flip the background spins during its motion is higher if the background spins are in the opposite direction [Fig. \ref{Fig2_3}(a)] than if those have the same polarity as the skyrmion [Fig. \ref{Fig2_3}(b)]. We have obtained analogous results for the CPP scenario.

\begin{figure}[!t]
\centering
\includegraphics[width=7cm]{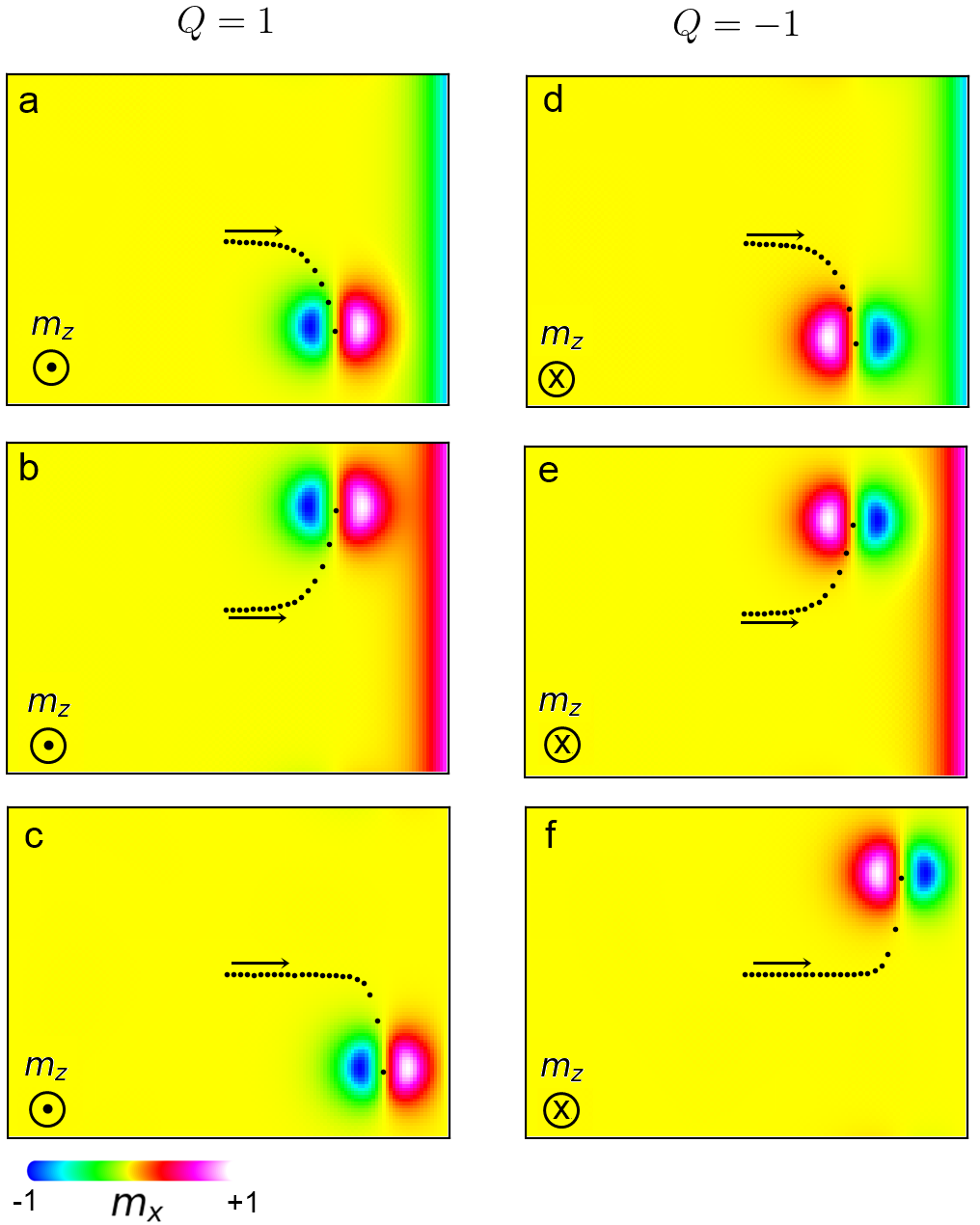}
\caption{Skyrmion trajectories (given as a trail of black dots) for $Q=1$ (a-c) and $Q=-1$ (d-f), for different canting of the magnetization at the edge, $m^x_{\text{fixed}}=-\sin\theta_{\text{edge}}$, $+\sin\theta_{\text{edge}}$ and $0$, respectively from top to bottom, with $\theta_{\text{edge}}=\frac{\pi}{4}$.}
\label{Fig1_3}
\end{figure}

In order to provide a better comparison of the simulations with the analytic results, we next consider the case $\alpha\ll 1$ and $\beta=0$. In this case, the relevant motion in the $y$ direction will be given solely by the effect of the external potential [see Eq. (\ref{eq.ymax})]. In Fig. \ref{Fig1_3} we take $\alpha=0.02$ (within the typical range $\alpha\sim10^{-3}\numrange{}{}10^{-2}$ for skyrmion-hosting materials \cite{Tomasello2014AMemories,ikeda2010perpendicular,Iwasaki2013UniversalMagnets}), and perform the same simulation of Fig. \ref{Fig2_3}, but now for six different situations: for the skyrmion numbers $Q=\pm1$, and the fixed magnetizations at the right edge $m^x_{\text{fixed}}=-\sin\theta_{\text{edge}}$, $+\sin\theta_{\text{edge}}$ and $0$, with $\theta_{\text{edge}}=\frac{\pi}{4}$. In the last case, the fixed spins do not induce any canting of the magnetization, but increase the necessary energy to flip their neighbours and must act as a repulsive potential for both considered skyrmion numbers. With such examples, we look for the corroboration of Eq. (\ref{Eq.x..}) for predicted deflection of the skyrmion trajectory. Comparing Fig. \ref{Fig1_3}(a,b) to (d,e), we change the skyrmion number [consequently the sign of $\mathcal{G}$ in Eq. (\ref{Eq.x..}) as well], but we also change the canting effect from repulsive to attractive and vice versa. Therefore, the skyrmion is deflected \textit{in the same direction} for both $Q=\pm1$. Comparing Fig. \ref{Fig1_3}(c) to (f), the skyrmion number changes, but the fixed spins act as a repulsive barrier in both situations, hence, the skyrmion is deflected \textit{in opposite directions} for opposite topological charge. These results are in complete accordance with Eq. (\ref{Eq.x..}) and will be useful to understanding the results of the next sections. For the CPP scenario, we obtained similar results when choosing $\alpha=0.02$.       

\subsubsection{Skyrmion trajectory while crossing an interface where DMI changes}

\begin{figure}[!t]
\centering
\includegraphics[width=7cm]{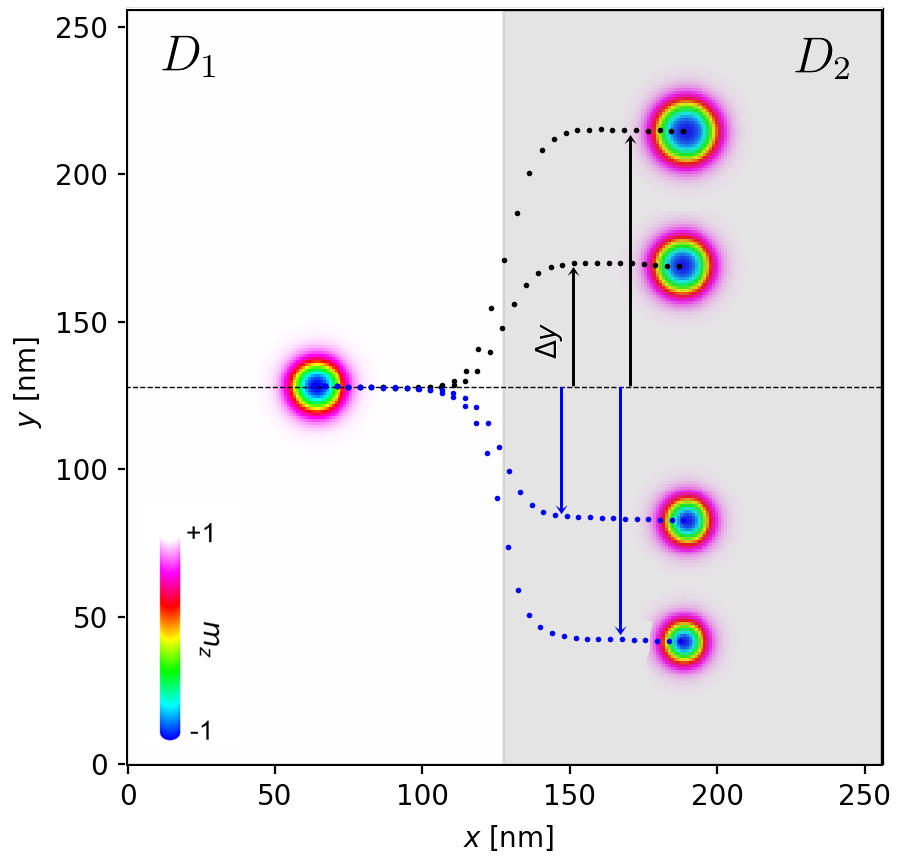}
\caption{The skyrmion is initialized on the left side of the diagram. Depending on the difference of the DMI strengths, $D_1$ and $D_2$, the skyrmion deflection is positive ($\Delta y>0$, black arrows) or negative ($\Delta y<0$, blue arrows). The dots show the respective trajectories of the skyrmion, for $j=20\times10^{10}$Am$^{-2}$ and  $\Delta D= 0.05,0.025,-0.025,-0.05$, respectively top-to-bottom. }
\label{Figr2}
\end{figure}

The examples of the previous subsection are not realistic, however, as we saw before, similar canting of the magnetization is intrinsic to the DMI interface(s) in a heterochiral ferromagnetic film. Therefore, we study next the trajectory of a single skyrmion while crossing an interface where DMI changes. In the simulations, we consider a sample of size $256\times256\times0.4$ nm$^3$, with DMI strength $D_1$ for $x<x_0$ and $D_2$ for $x>x_0$, where $x_0=128$ nm. The skyrmion is initialized at the position $x=64$ nm, $y=128$ nm (see Fig. \ref{Figr2}) and we consider periodic boundary conditions in the $y$ direction. An in-plane current is applied along $-\hat{x}$ (CIP scenario), with $\alpha=0.02$ and $\beta=0$, such that the relevant motion in the $y$ direction will be given solely by the effect of the heterochiral interface. As expected from the previous discussion, the skyrmion is deflected at the interface along $\pm \hat{y}$, depending on the canting direction, which in turn depends on the DMI strengths $D_1$ and $D_2$. 

\begin{figure}[!t]
\centering
\includegraphics[width=\linewidth]{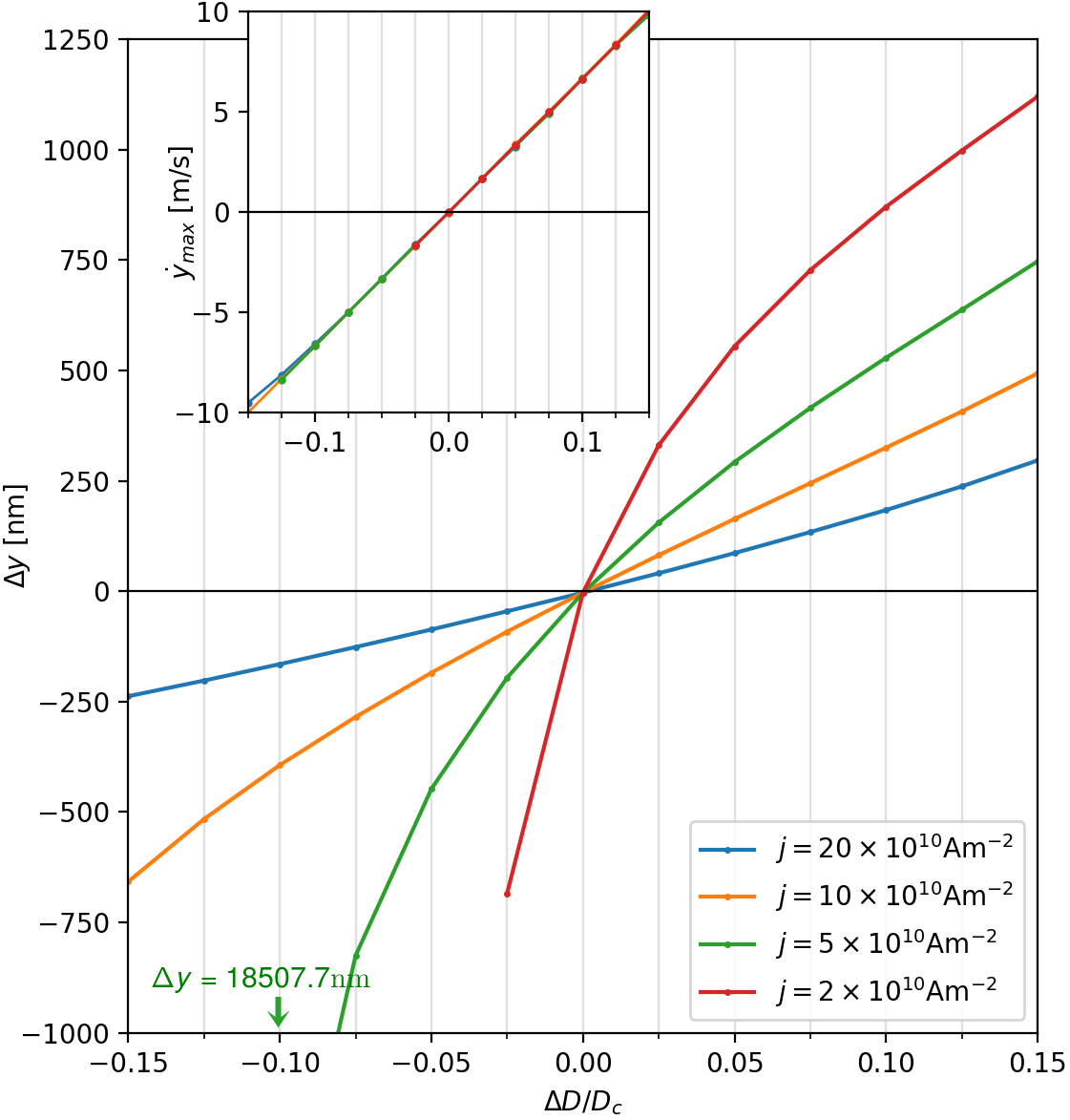}
\caption{Main panel exhibits the deflection in the skyrmion trajectory, $\Delta y$, for different values of applied current $j$ and the change in DMI across the interface $\Delta D=D_2-D_1$, with $D_1=0.8D_c$ fixed. The inset shows the maximal velocity of the skyrmion in the $y$ direction (along the interface) as a function of $\Delta D$, for different values of $j$.}
\label{Figr3}
\end{figure}

To illustrate the role of different parameters, we calculate the skyrmion deflection $\Delta y$ after the skyrmion reaches the position $x=192$ nm (as shown in Fig. \ref{Figr2}), for selected values of DMI strengths and applied currents. Fig. \ref{Figr3} shows the skyrmion deflection after crossing the interface as a function of $\Delta D=D_2-D_1$, with $D_1=0.8D_c$ fixed. For high currents, the deflections are smaller than those observed for low currents, and anti-symmetric for $\Delta D$ positive or negative, since the energy barrier induced by the interface is small when compared to the kinetic energy induced by the applied current. On the other hand, for low currents, the skyrmion motion can be completely blocked by the repulsive potential induced by $\Delta D<0$, if $j<j_c(\Delta D)$, where $j_c$ is the critical current for the skyrmion to overcome the interface. The more negative $\Delta D$ is, the higher is the necessary current for the skyrmion to overcome the interface. For example, for $j=2\times 10^{10}$ Am$^{-2}$ in Fig. \ref{Figr3}, the skyrmion can not cross the interface for $\Delta D \leq -0.05D_c$, and continues the motion purely in the $y$ direction, along the interface. Notice that for the considered parameters the $y$ component of the skyrmion velocity \textit{does not depend on the applied current} [see Eq. (\ref{eq.ymax})], as confirmed by the graph in the inset of Fig. \ref{Figr3}. The maximal velocity of the skyrmion in the $y$ direction, for a fixed $\Delta D$, is the same for all current values. However, for low currents the skyrmion takes longer time to cross the interface, which translates into a larger deflection. The largest deflections are observed when the applied current is just above $j_c$, e.g., for the case of $j=5\times 10^{10}$ Am$^{-2}$ in Fig. \ref{Figr3}, where the skyrmion overcomes the interface with $\Delta D=-0.1D_c$ after a rather extreme deflection of $\Delta y=-18.5$ $\mu$m. 

Comparing the graph in the inset of Fig. \ref{Figr3}, where $\Dot{y}_{\text{max}}$ varies linearly with $\Delta D$, with Eq. (\ref{eq.ymax}) and Eq. (\ref{eq.theta0}), we obtain
\begin{equation}
    \Fmax\approx c\mathcal{G}\Delta D=c\pi D_c\mathcal{G}\sin\theta_0,
    \label{eq.Fmax}
\end{equation}
where $c$ is the slope of $\Dot{y}_{\text{max}}(\Delta D)$ characteristic in the inset of Fig. \ref{Figr3}, and $\theta_0$ is the canting angle at the interface (from the graph, we obtained $c\approx67$). Note that, since the external potential $V$ does not depend on the applied current scenario and Eq. (\ref{eq.ymax}) is valid for both CIP and CPP scenarios, the graph in the inset of Fig. \ref{Figr3} and consequently Eq. (\ref{eq.Fmax}) are general results for a ferromagnetic skyrmion. Therefore, the critical current, given by Eqs. (\ref{jcCIP}) and (\ref{Eq.R10}), depends linearly on $\Delta D$ for both CIP and CPP scenarios. Note that the dissipative factor $\mathcal{D}$ in Eqs. (\ref{jcCIP}) and (\ref{Eq.R10}) depends on the skyrmion size, which in turn depends on the material parameters, e.g., the DMI strength. However, as will be shown in Fig. \ref{Figjc} of the next section, for the considered range of parameters, such linear dependence is preserved in the CPP scenario for different values of $\alpha$.    

\subsubsection{Multi-channel skyrmion bit sequencer}\label{Sec.IIIC}

Based on our findings, the heterochiral interface can be used to very precisely guide the skyrmion motion in a more complex circuitry, for example to selectively ``write'' skyrmions in one of multiple nanotracks, or to selectively direct a skyrmion to one of the many logical gates in a larger skyrmion microprocessor. We here exemplify such an application of a heterochiral interface, for the targetted manipulation of a chain of skyrmions by pulsed current. Although simplistic, this example is intended for the reader to creatively visualize other possible uses of heterochiral systems.
\begin{figure}[!t]
\centering
\includegraphics[width=\linewidth]{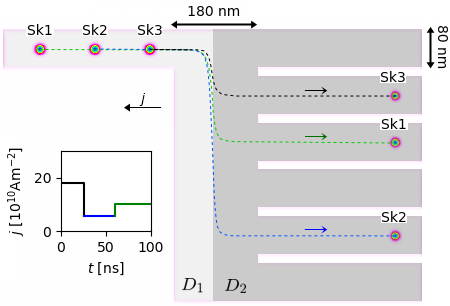}
\caption{Selective deflection of a skyrmion chain into multiple nanotracks, using the properties of a heterochiral interface. Dashed lines indicate the trajectory of each skyrmion during the simulation, for a series of current pulses of $j=18\times10^{10}$~Am$^{-2}$ for $0<t<25$~ns, $j=5.5\times10^{10}$~Am$^{-2}$ for $25<t<60$~ns, and $j=10\times10^{10}$~Am$^{-2}$ for $t>60$~ns, in a sample with $\alpha=0.02$, $\beta=0$, $D_1=0.8D_c$ and $D_2=0.75D_c$ ($\Delta D/D_c=-0.05$).} 
\label{Fig9}
\end{figure}

In this example, we consider a rectangular ferromagnetic film of size $880\times634\times0.4$ nm$^3$, where high-DMI tracks are engineered (by a suitable heavy-metal capping layer, see Fig. \ref{Fig9}) with DMI strengths of $D_1=0.8D_c$ (single track on the left) and $D_2=0.75D_c$ (six tracks on the right side). A skyrmion chain, containing skyrmions labeled Sk1, Sk2 and Sk3 and separated by $115$~nm, is initialized in the $D_1$ track on the left side of the sample. A current pulse is then applied along the $-\hat{x}$ direction (CIP scenario), which induces motion of skyrmions along $+\hat{x}$ direction. The duration and intensity of subsequent current pulses is designed in such a manner that each skyrmion reaches the heterochiral interface under a different current density, and thereby experiences different deflection of its trajectory. Moreover, the intensity of the pulses is chosen according to Fig. \ref{Figr3}, so that the deflection of the skyrmions exactly corresponds to the entry point of one of the six tracks on the right side of the sample~\cite{supplemental}.

Obviously, the exact duration and intensity of the current pulses has to be precisely engineered for a particular realization of the sample, depending on the separation of skyrmions in the initial chain and the values of all relevant parameters including the change of DMI across the interface. Nevertheless, once optimized, such an interface can be very reliably used to write skyrmions in multiple channels in any desired sequence, as we show in the animated data in Supplementary Material~\cite{supplemental}. We remind the reader that current-induced deflection of a FM skyrmion at a heterochiral interface can easily exceed ten micrometers (as shown in the previous section), hence a large number of nanotracks could be very controllably accessed in this manner. We stress that such controlled manipulation is needed for more complex skyrmion-based computing and storage circuits. For instance, it could be used to selectively place the skyrmions in the input branches of (many) skyrmion-based logic gates~\cite{zhang2015magnetic}, or to precisely write information in multi-bit memory cells. 

\subsection{Antiferromagnetic skyrmion}\label{Sec.IIIB}

Antiferromagnetic (AFM) skyrmions are expected to combine the advantages of antiferromagnets with those of skyrmions regarding spintronic applications. AFM skyrmions have zero net topological charge and simulations of their current-induced motion have shown that accordingly they move straight along the direction imposed by the applied current~\cite{Barker2016StaticTemperature,Zhang2016AntiferromagneticManipulation,Jin2016DynamicsEffect,Xia2017ControlRacetrack}. This is considered advantageous for applications, because as opposed to ferromagnetic skyrmions their antiferromagnetic counterparts are not driven towards the boundary of the hosting magnetic structures, where they can collapse. Additional benefits arise from their antiferromagnetic nature, i.e. their insensitivity to parasitic stray fields \cite{Barker2016StaticTemperature}. In what follows, we address in more detail the behavior of AFM skyrmions in heterochiral samples. 
\begin{figure}[!t]
\centering
\includegraphics[width=\linewidth]{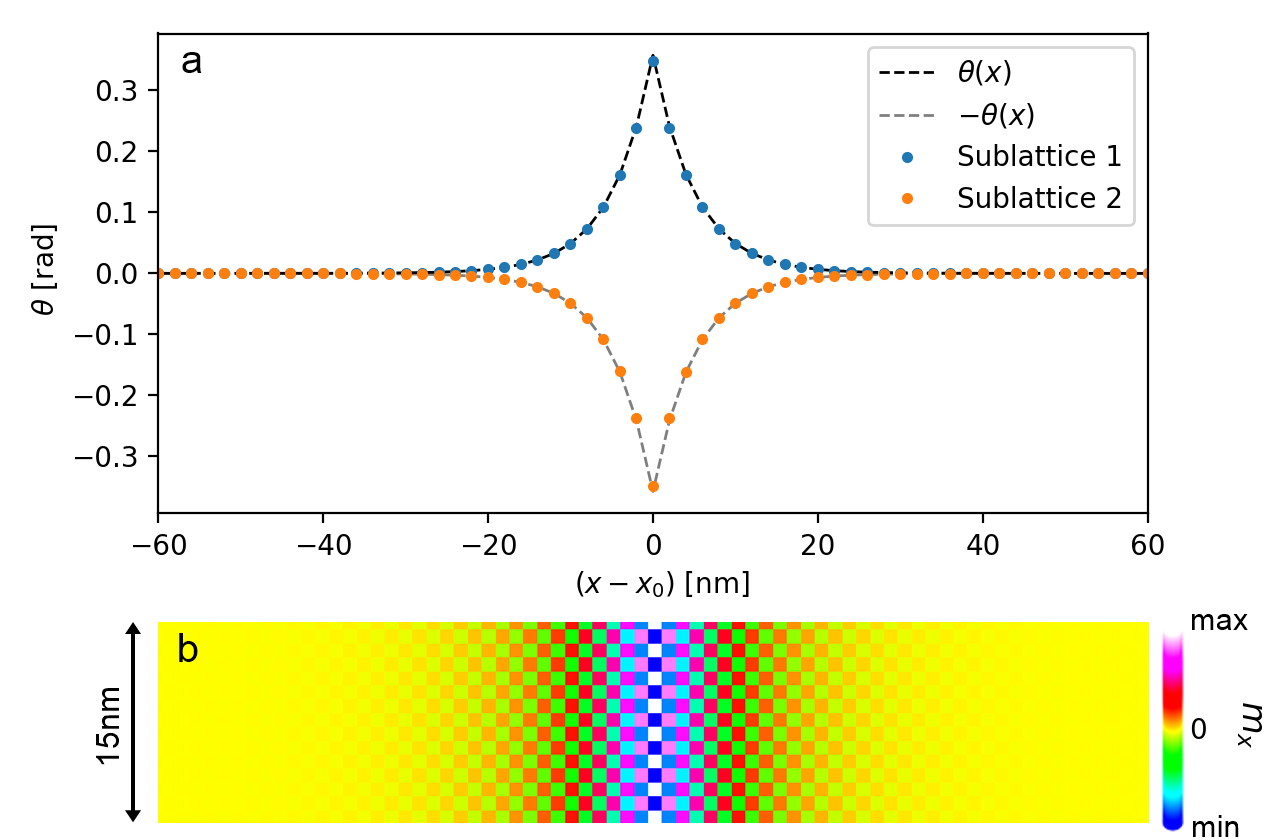}
\caption{(a) Canting of the magnetization $\theta(x)$ at a DMI interface of a heterochiral AFM system, plotted separately for each sublattice, for $D_2-D_1=1.1D_c$. Dashed lines represent the analytic result for the FM system, given by Eq. (\ref{eq.theta}). (b) Snapshot zoom of the  configuration obtained after minimizing the energy numerically.}
\label{Figr6} 
\end{figure}
\begin{figure}[!t]
\centering
\includegraphics[width=\linewidth]{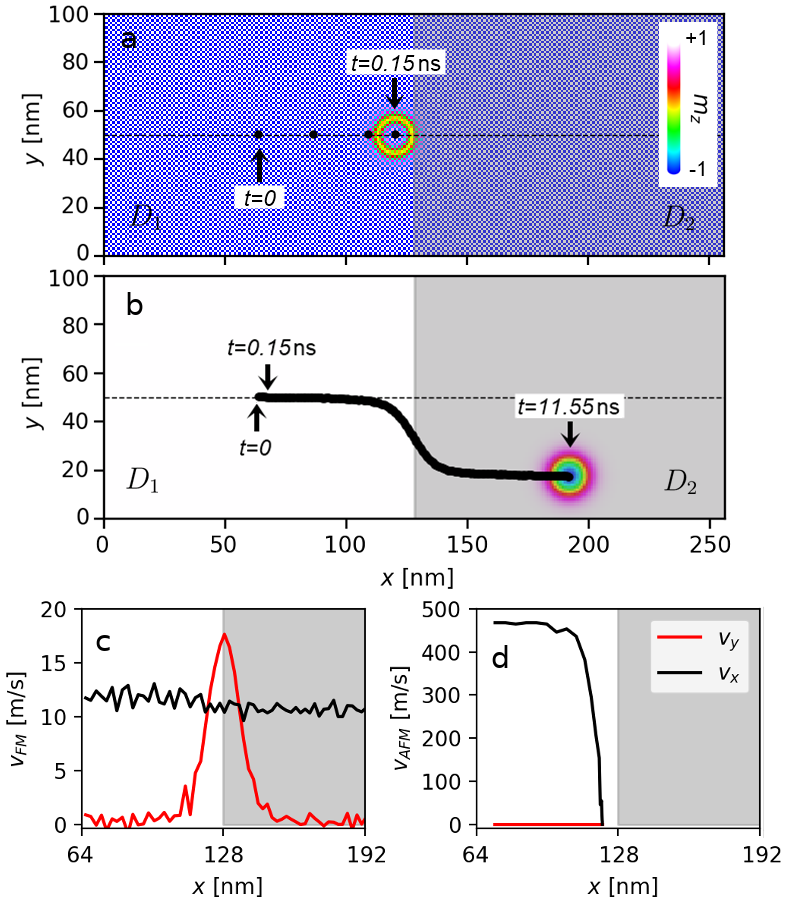}
\caption{Snapshots of the calculated spin configurations during simulation for the AFM (a) and FM (b) skyrmion driven by the CPP scenario. The trail of black dots indicates the skyrmion trajectory. (a) The AFM skyrmion reaches the interface after $t=0.15$ ns, where its movement is completely blocked. (b) The FM skyrmion moves slower than the AFM one, but can cross the interface after sufficiently long time. (c) Center-of-mass velocities of the FM skyrmion, during motion shown in (b). (d) Center-of-mass velocities of the AFM skyrmion, during motion shown in (a). The spin current is applied along the $\hat{z}$ direction but polarized along $+\hat{y}$ ($\hat{j}_{\text{hm}}=-\hat{x}$), for the AFM case, and $-\hat{x}$ ($\hat{j}_{\text{hm}}=-\hat{y}$) for the FM case, with current density $j=2\times10^{10}$ Am$^{-2}$, DMI strengths $D_1=0.8D_c$ and $D_2=0.775D_c$, and damping parameter $\alpha=0.02$.}
\label{Figr7}
\end{figure}
\begin{figure}[!t]
\centering
\includegraphics[width=\linewidth]{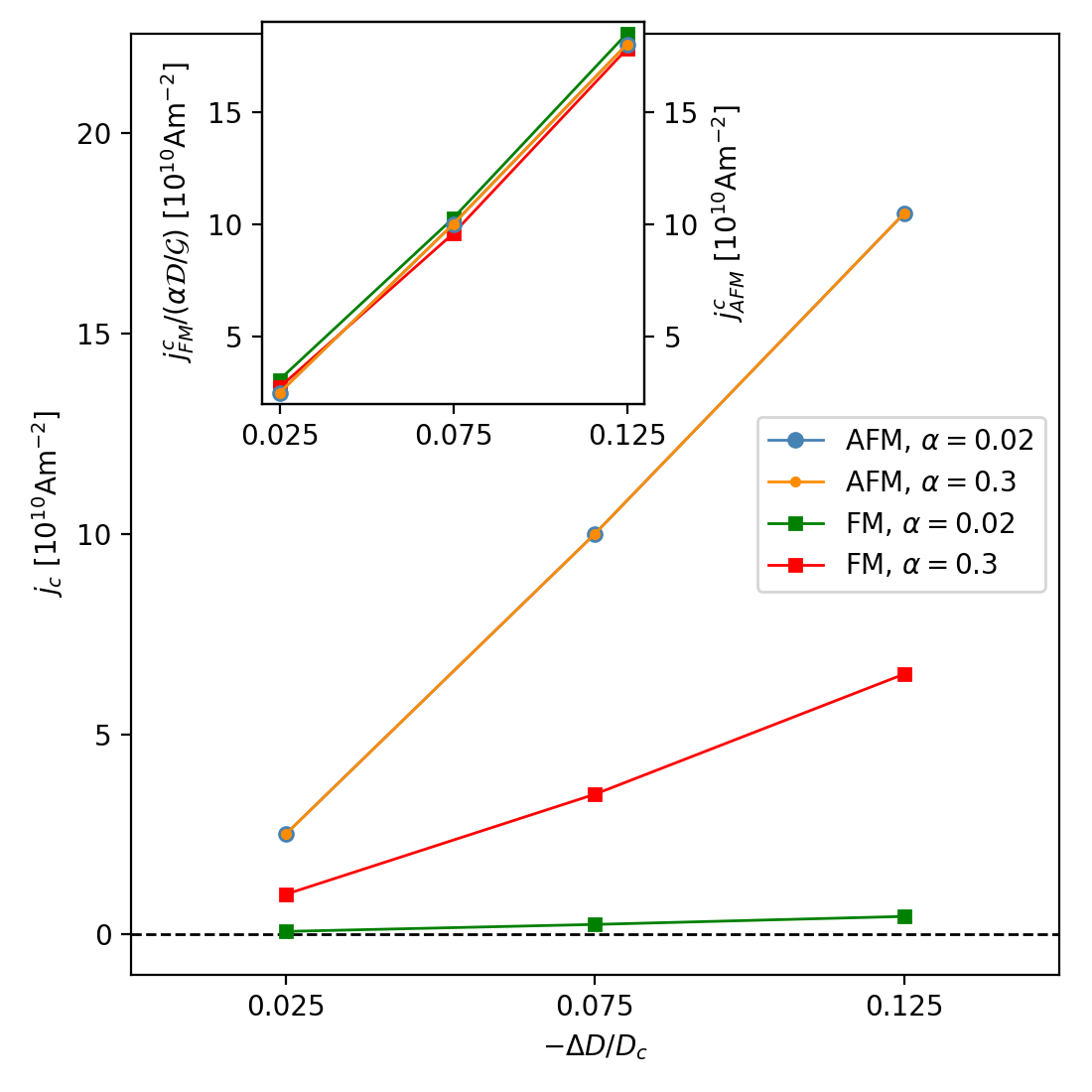}
\caption{Critical current for the skyrmion to overcome a heterochiral interface in the AFM and FM cases, as a function of $\Delta D$, with $D_1=0.8D_c$ fixed. Inset shows that all data collapses on the same curve with appropriate scaling, following Eq. (\ref{Eq.jcAFM}).}
\label{Figjc}
\end{figure}

Antiferromagnetic skyrmions have been recently intensively studied regarding their spin structure, their stability, and their motion \cite{Barker2016StaticTemperature,Zhang2016AntiferromagneticManipulation,Jin2016DynamicsEffect,Xia2017ControlRacetrack,rosales2015three,khoshlahni2018ultrafast,bessarab2017stability}. The AFM skyrmion comprises a two-sublattice structure, where each sublattice (indexed $1$ and $2$) contains half of the spins of the system and has the opposite magnetization of the other sublattice. In this way, the topological numbers projected to each sublattice satisfy $Q_1=-Q_2$. The opposing topological index of two sublattices causes the exact cancellation of the Magnus force in the presence of current, so the antiferromagnetic skyrmion moves along the direction of the current. The velocity of the AFM skyrmion driven by a current density is inversely proportional to the damping factor $\alpha$, and the AFM skyrmion can move much faster than the FM one for weak damping, possibly reaching km/s while remaining stable \cite{Barker2016StaticTemperature,Zhang2016AntiferromagneticManipulation,Jin2016DynamicsEffect,Xia2017ControlRacetrack,shen2018dynamics}.

To understand the dynamics of the AFM skyrmion while crossing an interface where the DMI changes, we first simulate, in the micromagnetic framework, the AFM ground state in the presence of such an interface. Here, we consider a sample of size $256\times100\times0.4$ nm$^3$ with DMI strength $D_1$ for $x<x_0$ and $D_2$ for $x>x_0$, with $x_0=128$nm. Fig.~\ref{Figr6}(b) shows a snapshot zoom of the  configuration obtained after minimizing the energy numerically. Notice that the induced canting ($m_x=\sin\theta$) points in opposite directions in each sublattice.  As shown in Fig.~\ref{Figr6}(a), the canting induced at each sublattice follows the analytic result for the FM system (dashed lines), given by Eq.~(\ref{eq.theta}).   

In the presence of canting induced by the DMI interface, we expect skyrmion scattering at each sublattice to follow the FM result of Fig.~\ref{Fig1_3}. We recall that the sublattices have opposite topological charge and induce opposite canting at the interface, hence the effective motion of the AFM skyrmion, given by a combination of the two sublattices, is a combination of either Fig.~\ref{Fig1_3} (a) and (e), or Fig.~\ref{Fig1_3} (b) and (d). Therefore, the characteristic deflection observed for the FM skyrmion while crossing the interface is completely absent (cancelled out) in the AFM system. However, the attractive or repulsive effect in the $x$ direction is still expected. 

The dynamics of the magnetization in the micromagnetic simulations is controlled by applying a spin current perpendicular to the plane (CPP scenario). Since the DMI interface is always either attractive in both sublattices [combination of Fig.~\ref{Fig1_3} (b) and (d)] or repulsive in both sublattices [combination of Fig.~\ref{Fig1_3} (a) and (e)], the DMI interface can be seen as an external potential in the modified Thiele equation [Eq.~(\ref{eq.CPP})], for a single lattice with $G=0$. Since  the AFM skyrmion moves along the direction of the current, now we assume $\textbf{j}_{\text{hm}}=j_{\text{hm}}\hat{x}$, and the Thiele equation for the AFM skyrmion reads
\begin{equation}
    -\alpha\mathcal{D}\Dot{x} +4\pi\mathcal{B}j_{\text{hm}}-\frac{dV}{dx}=0,\label{eq.thieleAFM}
\end{equation}
with $\Dot{y}=0$. In the same way as in the FM case, the critical current for the AFM skyrmion to overcome a repulsive potential is obtained by choosing $\Dot{x}=0$ in Eq. (\ref{eq.thieleAFM}), for the maximal value of $F=dV/dx$, i.e.
\begin{equation}
    4\pi\mathcal{B}j_{\text{hm}}^c=\Fmax,
    \label{Eq.jcAFM}
\end{equation}
which means that the critical current does not depend on $\alpha$ [similar result is obtained for the CIP scenario if one considers $\mathcal{G}=0$ in Eq. (\ref{eq.R1})]. Therefore, for $\alpha\ll1$, not only that AFM skyrmion travels faster than the corresponding FM one, but the critical current for the AFM skyrmion to overcome the same energy barrier is much higher than that expected for the FM skyrmion [see Eq. (\ref{Eq.R10})]: $j_{\text{AFM}}^c= (\mathcal{G}/\mathcal{D}\alpha)j_{\text{FM}}^c$. Fig. \ref{Figr7} shows the comparison between the AFM and FM skyrmion driven by the CPP scenario in the presence of a DMI interface. The skyrmion is initialized on the left side of the interface, and for the same current density ($j=2\times10^{10}$ Am$^{-2}$), the AFM skyrmion moves much faster than the FM one, as shown in Fig. \ref{Figr7}(c,d), but only the FM skyrmion can cross the interface. This means that the enhanced skyrmion confinement reported in ferromagnetic high-DMI racetracks due to spatially engineered DMI\cite{Mulkers2017EffectsFilms} is even more effective for the antiferromagnetic racetracks.

Fig. \ref{Figjc} shows the numerically calculated critical current for the skyrmion to overcome the heterochiral interface in the AFM and FM cases, as a function of the difference in DMI across the interface. For the FM case, for lowered damping parameter, the skyrmion moves faster, but the efficiency of the confinement also decreases, which may be a drawback for racetrack applications. However, in the AFM case, the critical current is very large for the considered values of $\alpha$, as expected from Eq. (\ref{Eq.jcAFM}). In other words, our results indicate that the AFM skyrmion indeed moves \textit{faster}, yet experiences \textit{stronger} confinement than the FM skyrmion in heterochiral films, especially for systems with weak damping. Both these (seemingly contradictory) features establish AFM skyrmions as a favorable choice for skyrmion-based devices. The values of $\alpha$ considered here are similar to those obtained from experimental results on CoFeB/W systems \cite{pai2012spin,ikeda2010perpendicular,Tomasello2014AMemories} and  Co/Pt layers \cite{metaxas2007creep} ($\alpha\approx0.015$ and $\alpha\approx0.3$, respectively). The inset in Fig. \ref{Figjc} demonstrates the scaling of the critical current of the FM cases to the AFM results, with factor $(\mathcal{D}\alpha/\mathcal{G})$, as expected from the analytic formulae. Here $\mathcal{G}=4\pi$ and the dissipative term is calculated from the simulations as follows \cite{Iwasaki2013UniversalMagnets}
\begin{equation}
    \mathcal{D}=\pi\sum_{i=1}^{N}\left[\left(\frac{\theta(i+1)-\theta(i-1)}{2} \right)^2 + \frac{\sin^2\theta(i)}{i^2} \right],
    \label{Eq.D}
\end{equation}
where $r=ia$ is the distance from the skyrmion core ($a$ is the lattice constant) and $\cos\theta=m_z$. For the inset in Fig. \ref{Figjc} we use $\mathcal{D}=4.87\pi$, calculated for the skyrmion at rest, in the region with $D_1=0.8D_c$. Note that similar results can be obtained from the Thiele equation by considering the external potential due to a ``boundary'' instead of the heterochiral interface (as done in Sec. \ref{Sec.IIIA1}).

\section{Conclusions}\label{Sec.IV}

Recent advances in atomically controlled growth of heterostructures have opened the door to heterochiral structures with spatially engineered DMI, with precisely defined interfaces where DMI changes. In this paper, we have addressed the expected behavior of skyrmions in such systems, by studying the dynamics of both ferromagnetic and antiferromagnetic skyrmions when encountering a heterochiral interface during their motion. We demonstrated that a local canting of the magnetization, characteristic for the interface where the DMI changes, can strongly deflect the trajectory of a FM skyrmion. We explored the thresholds of this phenomenon both analytically and numerically, and quantified its dependence on the relevant material parameters. These findings are very useful for the controlled manipulation of either single skyrmions or skyrmion chains in skyrmion-based devices (switches, logic gates, memory elements, to name a few) where depending on the applied current one can control which path the skyrmion will take in the corresponding nanoengineered circuit, as exemplified in Sec.~\ref{Sec.IIIC}.

In addition, we showed that such a deflection characteristic for the ferromagnetic skyrmion is completely absent in the antiferromagnetic case. Although this finding is detrimental for the applications of the above effect in AFM systems, we demonstrated that the AFM skyrmion holds other advantages to the FM one - it travels much faster for the given applied current, yet is far better confined in heterochiral films even at high driving currents. This makes AFM skyrmions favorable for skyrmion-based devices in which very fast transfer of information and reliable guidance within specified tracks are essential.

\section*{Acknowledgements}
This work was supported by the Research Foundation - Flanders (FWO-Vlaanderen) and  Brazilian Agencies FACEPE, under the grant No. APQ-0198-1.05/14, CAPES 
and CNPq.

\bibliography{references}

\end{document}